\documentclass[12pt]{article}

\usepackage{times}

\usepackage[pdftex]{graphicx} 

\usepackage{amssymb, amsmath}

\newenvironment{abs}{%
\begin{quote} \bf}
{\end{quote}}

\title{Propagation of an Earth-directed coronal mass ejection in three dimensions} 

\author
{Jason P. Byrne$^{1}$, Shane A. Maloney$^{1}$, R. T. James McAteer$^{1}$, \\ Jose M. Refojo$^{2}$ and Peter T. Gallagher$^{1}$$^{\ast}$\\
\\
\normalsize{$^{1}$Astrophysics Research Group, School of Physics, Trinity College Dublin, Dublin 2, Ireland.}\\
\normalsize{$^{2}$Trinity Centre for High Performance Computing, Trinity College Dublin, Dublin 2, Ireland.}\\
\\
\normalsize{$^\ast$Correspondence to: Peter T. Gallagher (peter.gallagher@tcd.ie)}
}

\date{}

\usepackage{color}

\begin{document} 

\baselineskip24pt

\maketitle

\begin{abs}

Solar coronal mass ejections (CMEs) are the most significant drivers of adverse space weather at Earth, but the physics governing their propagation through the heliosphere is not well understood. While stereoscopic imaging of CMEs with the Solar Terrestrial Relations Observatory (STEREO) has provided some insight into their three-dimensional (3D) propagation, the mechanisms governing their evolution remain unclear due to difficulties in reconstructing their true 3D structure. Here we use a new elliptical tie-pointing technique to reconstruct a full CME front in 3D, enabling us to quantify its deflected trajectory from high latitudes along the ecliptic, and measure its increasing angular width and propagation from 2\,--\,46~R$_{\odot}$ ($\sim$0.2~AU). Beyond 7~R$_{\odot}$, we show that its motion is determined by aerodynamic drag in the solar wind and, using our reconstruction as input for a 3D magnetohydrodynamic simulation, we determine an accurate arrival time at the L1 point near Earth.

\end{abs}

\subsection*{}

CMEs are spectacular eruptions of plasma and magnetic field from the surface of the Sun into the heliosphere. Travelling at speeds of up to 2,500~km~s$^{-1}$ and with masses of up to 10$^{16}$~g, they are recognised as drivers of geomagnetic disturbances and adverse space weather at Earth and other planets in the solar system$^{1,2}$. Impacting our magnetosphere with average magnetic field strengths of 13~nT and energies of $\sim$10$^{25}$~J they can cause telecommunication and GPS errors, power grid failures, and increased radiation risks to astronauts$^{3}$. It is therefore important to understand the forces that determine their evolution, in order to better forecast their arrival time and impact at Earth and throughout the heliosphere.

Identifying the specific processes that trigger the eruption of CMEs is the subject of much debate, and many different models exist to explain these$^{4-6}$. One common feature is that magnetic reconnection is responsible for the destabilisation of magnetic flux ropes on the Sun, which then erupt through the corona into the solar wind to form CMEs$^{7}$. In the low solar atmosphere, it is postulated that high-latitude CMEs undergo deflection since they are often observed at different position angles than their associated source region locations$^{8}$. It has been suggested that field lines from polar coronal holes may guide high-latitude CMEs towards the equator$^{9}$, or that the initial magnetic polarity of a flux rope relative to the background magnetic field influences its trajectory$^{10,11}$. During this early phase, CMEs are observed to expand outwards from their launch site, though plane-of-sky measurements of their increasing sizes and angular widths are ambiguous in this regard$^{12}$. This expansion has been modelled as being due to a pressure gradient between the flux rope and the background solar wind$^{13,14}$. At larger distances in their propagation, CMEs are expected to interact with the solar wind and the interplanetary magnetic field. Studies that compare in-situ CME velocity measurements with initial eruption speeds through the corona show that slow CMEs are accelerated toward the speed of the solar wind, and fast CMEs decelerated$^{15,16}$. It has been suggested that this is due to the effects of drag acting on the CME in the solar wind$^{17,18}$. However, the quantification of drag along with that of both CME expansion and non-radial motion is currently lacking, due primarily to the limits of observations from single fixed viewpoints with restricted fields-of-view. The projected 2D nature of these images introduces uncertainties in kinematical and morphological analyses, and therefore the true 3D geometry and dynamics of CMEs has been difficult to resolve. Efforts were made to infer 3D structure from 2D images recorded by the Large Angle Spectrometric Coronagraph (LASCO) on board the Solar and Heliospheric Observatory (SOHO), situated at the first Lagrangian L1 point. These efforts were based upon either a pre-assumed geometry of the CME$^{19,20,21}$ or a comparison of observations with in-situ and on-disk data$^{22,23}$. Of note is the polarisation technique used to reconstruct the 3D geometry of CMEs in LASCO data$^{24}$, though this is only valid for heights of up to 5~R$_{\odot}$ (1~R$_{\odot}$~=~6.95$\times$10$^{5}$~km).

Recently, new methods to track CMEs in 3D have been developed for the STEREO mission$^{25}$. Launched in 2006, STEREO comprises of two near-identical spacecraft in heliocentric orbits ahead and behind the Earth, which drift away from the Sun-Earth line at a rate of $\pm$22$^{\circ}$ per year. This provides a unique twin perspective of the Sun and inner heliosphere, and enables the implementation of a variety of methods for studying CMEs in 3D$^{26}$. Many of these techniques are applied within the context of an epipolar geometry$^{27}$. One such technique consists of tie-pointing lines-of-sight across epipolar planes, and is best for resolving a single feature such as a coronal loop on-disk$^{28}$. Under the assumption that the same feature may be tracked in coronagraph images, many CME studies have also employed tie-pointing techniques with the COR1 and COR2 coronagraphs of the Sun-Earth Connection Coronal and Heliospheric Investigation (SECCHI$^{29}$) aboard STEREO$^{30-32}$. The additional use of SECCHI's Heliospheric Imagers (HI1/2) allows a study of CMEs out to distances of 1 astronomical unit (1~AU~=~149.6$\times$10$^{6}$~km), however a 3D analysis can only be carried out if the CME propagates along a trajectory between the two spacecraft so that it is observed by both HI instruments. Otherwise, assumptions of its trajectory have to be inferred from either its association with a source region on-disk$^{33}$ or its trajectory through the COR data$^{15}$, or derived by assuming a constant velocity through the HI fields-of-view$^{34}$. Triangulation of CME features using time-stacked intensity slices at a fixed latitude, named `J-maps' due to the characteristic propagation signature of a CME, has also been developed$^{35,36}$. This technique is hindered by the same limitation of standard tie-pointing techniques; namely that the curvature of the feature is not considered, and the intersection of sight-lines may not occur upon the surface of the observed feature. Alternatively, forward modelling of a 3D flux rope based upon a graduated cylinder model may be applied to STEREO observations$^{37}$. Some of the parameters governing the model shape and orientation may be changed by the user to best fit the twin observations simultaneously, though the assumed flux rope geometry is not always appropriate. We have developed a new 3D triangulation technique that overcomes the limitations of previous methods by considering the curvature of the CME front in the data. This acts as a necessary third constraint on the reconstruction of the CME front from the combined observations of the twin STEREO spacecraft. Applying this to every image in the sequence enables us to investigate the changing dynamics and morphology of the CME as it propagates from the Sun into interplanetary space.

\subsection*{Results}

On 12 Dec. 2008 an erupting prominence was observed by STEREO while the spacecraft were in near quadrature at 86.7$^{\circ}$ separation (Fig.~1a). The eruption is visible at 50\,--\,55$^{\circ}$ north from 03:00 UT in SECCHI/Extreme Ultraviolet Imager (EUVI) images, obtained in the 304~\AA\ passband, in the north-east from the perspective of STEREO-(A)head and off the north-west limb from STEREO-(B)ehind. The prominence is considered to be the inner material of the CME which was first observed in COR1-B at 05:35~UT (Fig.~1b). For our analysis, we use the two coronagraphs (COR1/2) and the inner Heliospheric Imagers (HI1) (Fig.~1c). We characterise the propagation of the CME across the plane-of-sky by fitting an ellipse to the front of the CME in each image$^{38}$ (Supplementary Movie 1). This ellipse fitting is applied to the leading edges of the CME but equal weight is given to the CME flank edges as they enter the field-of-view of each instrument. The 3D reconstruction is then performed using a method of elliptical tie-pointing within epipolar planes containing the two STEREO spacecraft, illustrated in Fig.~2 (see Methods).
\newline
\newline
\textbf{Non-radial CME motion.} It is immediately evident from the reconstruction in Fig.~2c (and Supplementary Movie 2) that the CME propagates non-radially away from the Sun. The CME flanks change from an initial latitude span of 16\,--\,46$^{\circ}$ to finally span approximately $\pm$\,30$^{\circ}$ of the ecliptic (Fig.~3b). The mean declination, $\theta$, of the CME is well fitted by a power-law of the form $\theta(r)=\theta_{0}r^{-0.92}~(2~$R$_{\odot}<r<46~$R$_{\odot})$ as a result of this non-radial propagation. Tie-pointing the prominence apex and fitting a power-law to its declination angle results in $\theta^{prom}(r)=\theta_{0}^{prom}r^{-0.82}~(1~$R$_{\odot}<r<3~$R$_{\odot})$, implying a source latitude of $\theta_{0}^{prom}$(1~R$_{\odot}$)~$\approx$~54$^{\circ}$~N in agreement with EUVI observations. Previous statistics on CME position angles have shown that, during solar minimum, they tend to be offset closer to the equator as compared to those of the associated prominence eruption$^{39}$. The non-radial motion we quantify here may be evidence of the drawn-out magnetic dipole field of the Sun, an effect predicted at solar minimum due to the influence of the solar wind pressure$^{40,41}$. Other possible influences include changes to the internal current of the magnetic flux rope$^{11}$, or the orientation of the magnetic flux rope with respect to the background field$^{10}$, whereby magnetic pressure can act asymmetrically to deflect the flux rope pole-ward or equator-ward depending on the field configurations.
\newline
\newline
\textbf{CME angular width expansion.} Over the height range 2\,--\,46~R$_{\odot}$ the CME angular width ($\Delta\theta=\theta_{max}-\theta_{min}$) increases from $\sim$30$^{\circ}$ to $\sim$60$^{\circ}$ with a power-law of the form $\Delta\theta(r)=\Delta\theta_{0}r^{0.22}$~$(2~$R$_{\odot}<r<46~$R$_{\odot})$ (Fig.~3c). This angular expansion is evidence for an initial overpressure of the CME relative to the surrounding corona (coincident with its early acceleration inset in Fig.~3a). The expansion then tends to a constant during the later drag phase of CME propagation, as it expands to maintain pressure balance with heliocentric distance. It is theorised that the expansion may be attributed to two types of kinematic evolution, namely spherical expansion due to simple convection with the ambient solar wind in a diverging geometry, and expansion due to a pressure gradient between the flux rope and solar wind$^{13}$. It is also noted that the southern portions of the CME manifest the bulk of this expansion below the ecliptic (best observed by comparing the relatively constant `Midtop of Front' measurements with the more consistently decreasing `Midbottom of Front' measurements in Fig.~3b). Inspection of a Wang-Sheeley-Arge (WSA) solar wind model run$^{42}$ reveals higher speed solar wind flows ($\sim$650~km~s$^{-1}$) emanating from open-field regions at high/low latitudes (approximately 30$^{\circ}$ north/south of the solar equator). Once the initial prominence/CME eruption occurs and is deflected into a non-radial trajectory, it undergoes asymmetric expansion in the solar wind. It is prevented from expanding upwards into the open-field high-speed stream at higher latitudes, and the high internal pressure of the CME relative to the slower solar wind near the ecliptic accounts for its expansion predominantly to the south. In addition, the northern portions of the CME attain greater distances from the Sun than the southern portions as a result of this propagation in varying solar wind speeds, an effect predicted to occur in previous hydrodynamic models$^{14}$.
\newline
\newline
\textbf{CME drag in the inner heliosphere.} Investigating the midpoint kinematics of the CME front, we find the velocity profile increases from approximately 100\,--\,300~km~s$^{-1}$ over the first 2\,--\,5~R$_{\odot}$, before rising more gradually to a scatter between 400\,--\,550~km~s$^{-1}$ as it propagates outward (Fig.~3a). The acceleration peaks at approximately 100~m~s$^{-2}$ at a height of $\sim$3~R$_{\odot}$, then decreases to scatter about zero. This early phase is generally attributed to the Lorentz force whereby the dominant outward magnetic pressure overcomes the internal and/or external magnetic field tension. The subsequent increase in velocity, at heights above $\sim$7~R$_{\odot}$ for this event, is predicted by theory to result from the effects of drag$^{17}$, as the CME is influenced by the solar wind flows of $\sim$550~km~s$^{-1}$ emanating from latitudes $\gtrsim$\,$\pm$\,5$^{\circ}$ of the ecliptic (again from inspection of the WSA model). At large distances from the Sun, during this postulated drag-dominated epoch of CME propagation, the equation of motion can be cast in the form:
\begin{eqnarray}
	\label{drag}
	M_{cme} \frac{d v_{cme}}{d t}&=&-\frac{1}{2} \rho_{sw} ( v_{cme} - v_{sw} ) |  v_{cme} - v_{sw} |  A_{cme}  C_{D}
\end{eqnarray}
 This describes a CME of velocity $v_{cme}$, mass $M_{cme}$, and cross-sectional area $A_{cme}$ propagating through a solar wind flow of velocity $v_{sw}$ and density $\rho_{sw}$. The drag coefficient, $C_D$, is found to be of the order of unity for typical CME geometries$^{18}$, while the density and area are expected to vary as power-law functions of distance $r$. Thus, we parameterise the density and geometric variation of the CME and solar wind using a power-law$^{43}$ to obtain:
\begin{eqnarray}
         \label{pdrag}
	\frac{d v_{cme}}{d r} &=& -\alpha r^{-\beta} \frac{1}{v_{cme}}\left ( v_{sw} - v_{cme} \right )^\gamma
\end{eqnarray}
where $\gamma$ describes the drag regime, which can be either viscous ($\gamma$~=~1) or aerodynamic ($\gamma$~=~2), and $\alpha$ and $\beta$ are constants primarily related to the cross-sectional area of the CME and the density ratio of the solar wind flow to the CME ($\rho_{sw}/\rho_{cme}$). The solar wind velocity is estimated from an empirical model$^{44}$. We determine a theoretical estimate of the CME velocity as a function of distance by numerically integrating equation~(\ref{pdrag}) using a 4th order Runge-Kutta scheme and fitting the result to the observed velocities from $\sim$7\,--\,46~R$_{\odot}$. The initial CME height, CME velocity, asymptotic solar wind speed, and $\alpha$, $\beta$, and $\gamma$ are obtained from a bootstrapping procedure which provides a final best-fit to the observations and confidence intervals for the parameters (see Methods). Best-fit values for $\alpha$ and $\beta$ were found to be (4.55$^{+2.30}_{-3.27}$)$\times$10$^{-5}$ and -2.02$^{+1.21}_{-0.95}$ which agree with values found in previous modelling work$^{44}$. The best-fit value for the exponent of the velocity difference between the CME and the solar wind, $\gamma$, was found to be 2.27$^{+0.23}_{-0.30}$, which is clear evidence that aerodynamic drag ($\gamma$~=~2) acts during the propagation of the CME in interplanetary space.

The drag model provides an asymptotic CME velocity of 555$_{-42}^{+114}$~km~s$^{-1}$ when extrapolated to 1~AU, which predicts the CME to arrive one day before the Advanced Composition Explorer (ACE) or WIND spacecraft detect it at the L1 point. We investigate this discrepancy by using our 3D reconstruction to simulate the continued propagation of the CME from the Alfv\'{e}n radius ($\sim$21.5~R$_{\odot}$) to Earth using the ENLIL with Cone Model$^{21}$ at NASA's Community Coordinated Modeling Center. ENLIL is a time-dependent 3D magnetohydrodynamic (MHD) code that models CME propagation through interplanetary space. We use the height, velocity, and width from our 3D reconstruction as initial conditions for the simulation, and find that the CME is actually slowed to $\sim$342~km~s$^{-1}$ at 1~AU. This is as a result of its interaction with an upstream, slow-speed, solar wind flow at distances beyond 50~R$_{\odot}$.  This CME velocity is consistent with in-situ measurements of solar wind speed ($\sim$330~km~s$^{-1}$) from the ACE and WIND spacecraft at L1. Tracking the peak density of the CME front from the ENLIL simulation gives an arrival time at L1 of $\sim$08:09~UT on 16 Dec. 2008. Accounting for the offset in CME front heights between our 3D reconstruction and ENLIL simulation at distances of $21.5~$R$_{\odot}<r<46~$R$_{\odot}$ gives an arrival time in the range 08:09\,--\,13:20~UT on 16 Dec. 2008. This prediction interval agrees well with the earliest derived arrival times of the CME front plasma pileup ahead of the magnetic cloud flux rope from the in-situ data of both ACE and WIND (Fig.~4) before its subsequent impact at Earth$^{34,36}$.

\subsection*{Discussion}

Since its launch, the dynamic twin-viewpoints of STEREO have enabled studies of the true propagation of CMEs in 3D space. Our new elliptical tie-pointing technique uses the curvature of the CME front as a necessary third constraint on the two viewpoints to build an optimum 3D reconstruction of the front. Here the technique is applied to an Earth-directed CME, to reveal numerous forces at play throughout its propagation.

The early acceleration phase results from the rapid release of energy when the CME dynamics are dominated by outward magnetic and gas pressure forces. Different models can reproduce the early acceleration profiles of CME observations though it is difficult to distinguish between them with absolute certainty$^{45,46}$. For this event the acceleration phase coincides with a strong angular expansion of the CME in the low corona, which tends toward a constant in the later observed propagation in the solar wind. While, statistically, expansion of CMEs is a common occurrence$^{47}$, it is difficult to accurately determine the magnitude and rate of expansion across the 2D plane-of-sky images for individual events. Some studies of these single-viewpoint images of CMEs use characterisations such as the cone model$^{20,21}$ but assume the angular width to be constant (rigid cone) which is not always true early in the events$^{12,38}$. Our 3D front reconstruction overcomes the difficulties in distinguishing expansion from image projection effects, and we show that early in this event there is a non-constant, power-law, angular expansion of the CME. Theoretical models of CME expansion generally reproduce constant radial expansion, based on the suspected magnetic and gas pressure gradients between the erupting flux rope and the ambient corona and solar wind$^{14,48,49}$. To account for the angular expansion of the CME, a combination of internal overpressure relative to external gas and magnetic pressure drop-offs, along with convective evolution of the CME in the diverging solar wind geometry, must be considered$^{13}$.

During this early phase evolution the CME is deflected from a high-latitude source region into a non-radial trajectory as indicated by the changing inclination angle (Fig.~3b). While projection effects again hinder interpretations of CME position angles in single images, statistical studies show that, relative to their source region locations, CMEs have a tendency to deflect toward lower latitudes during solar minimum$^{39,50}$. It has been suggested that this results from the guiding of CMEs towards the equator by either the magnetic fields emanating from polar coronal holes$^{8,9}$ or the flow pattern of the background coronal magnetic field and solar wind/streamer influences$^{19,51}$. Other models show that the internal configuration of the erupting flux rope can have an important effect on its propagation through the corona. The orientation of the flux rope, either normal or inverse polarity, will determine where magnetic reconnection is more likely to occur, and therefore change the magnetic configuration of the system to guide the CME either equator- or pole-ward$^{10}$. Alternatively, modelling the filament as a toroidal flux rope located above a mid-latitude polarity inversion line results in non-radial motion and acceleration of the filament, due to the guiding action of the coronal magnetic field on the current motion$^{11}$. Both of these models have a dependence on the chosen background magnetic field configuration, and so the suspected drawn-out magnetic dipole field of the Sun by the solar wind$^{40,41}$ may be the dominant factor in deflecting the prominence/CME eruption into this observed non-radial trajectory.

At larger distances from the Sun ($>$\,7~R$_{\odot}$) the effects of drag become important as the CME velocity approaches that of the solar wind. The interaction between the moving magnetic flux rope and the ambient solar wind has been suggested to play a key role in CME propagation at large distances where the Lorentz driving force and the effects of gravity become negligible$^{4}$. Comparisons of initial CME speeds and in-situ detections of arrival times have shown that velocities converge on the solar wind speed$^{15,16}$. For this event we find that the drag force is indeed sufficient to accelerate the CME to the solar wind speed, and quantify that the kinematics are consistent with the quadratic regime of aerodynamic drag (turbulent, as opposed to viscous, effects dominate). The importance of drag becomes further apparent through the CME interaction with a slow-speed solar wind stream ahead of it, slowing it to a speed that accounts for the observed arrival time at L1 near Earth. This agrees with the conjecture that Sun-Earth transit time is more closely related to the solar wind speed than the initial CME speed$^{52}$. Other kinematic studies of this CME through the HI fields-of-view quote velocities of 411\,$\pm$\,23~km~s$^{-1}$ (Ahead) and 417\,$\pm$\,15~km~s$^{-1}$ (Behind) when assumed to have zero acceleration during this late phase of propagation$^{34}$, or an average of 363\,$\pm$\,43~km~s$^{-1}$ when triangulated in time-elongation J-maps$^{36}$. These speeds through the HI fields-of-view, lower than those quantified through the COR1/2 fields-of-view, agree somewhat with the deceleration of the CME to match the slow-speed solar wind ahead of it in our MHD simulation. Ultimately we are able to predict a more accurate arrival time of the CME front at L1.

A cohesive physical picture for how the CME erupts, propagates, and expands in the solar atmosphere remains to be fully developed and understood from a theoretical perspective. Realistic MHD models of the Sun's global magnetic field and solar wind are required to explain all processes at play, along with a need for adequate models of the complex flux rope geometries within CMEs. Additionally, ambitious space exploration missions, such as Solar Orbiter$^{53}$ (ESA) and Solar Probe$+$$^{54}$ (NASA), will be required to give us a better understanding of the fundamental plasma processes responsible for driving CMEs and determining their adverse effects at Earth. 

\subsection*{Methods}

\textbf{CME front detection and characterisation.} For the coronagraph images of COR1/2 a multiscale filter was used to determine a scale at which the signal-to-noise ratio of the CME was deemed optimal for the pixel-chaining algorithm to highlight the edges in the images$^{55 }$. In order to specifically determine the CME front, running and fixed difference masks were overlaid on the multiscale edge detections of both the Ahead and Behind viewpoints simultaneously, enabling us to confidently point-and-click along the relevant CME front edges in each image. For the Heliospheric images of HI1 a modified running difference was used to enhance the faint CME features by correcting for the apparent background stellar motion between frames$^{15}$. The CME was scaled to an appropriate level for point-and-clicking along its front. Once the CME fronts were determined across each instrument plane-of-sky, an ellipse was fit to each front in order to characterise the changing morphology of the CME$^{38}$.
\newline
\newline
\textbf{Elliptical tie-pointing.} 3D information may be gleaned from two independent viewpoints of a feature using tie-pointing techniques to triangulate lines-of-sight in space$^{27}$. However, when the object is known to be a curved surface, sight-lines will be tangent to it and not necessarily intersect upon it. Consequently CMEs cannot be reconstructed by tie-pointing alone, but rather their localisation may be constrained by intersecting sight-lines tangent to the leading edges of a CME$^{56,57}$. It is possible to extract the intersection of a given epipolar plane through the ellipse fits in both the Ahead and Behind images, resulting in a quadrilateral in 3D space. Inscribing an ellipse within the quadrilateral such that it is tangent to all four sides$^{58,59}$ provides a slice through the CME that matches the observations from each spacecraft. A full reconstruction is achieved by stacking ellipses from numerous epipolar slices. Since the positions and curvatures of these inscribed ellipses are constrained by the characterised curvature of the CME front in the stereoscopic image pair, the modelled CME front is considered an optimum reconstruction of the true CME front. This is repeated for every frame of the eruption to build the reconstruction as a function of time and view the changes to the CME front as it propagates in 3D.
\newline
\newline
Following Horwitz$^{59}$, we inscribe an ellipse within a quadrilateral using the following steps (see Fig.~5):
\begin{enumerate}
\item Apply an isometry to the plane such that the quadrilateral has vertices $(0,0)$, $(A,B)$, $(0,C)$, $(s,t)$, where in the case of an affine transformation we set $A=1$, $B=0$ and $C=1$, with $s$ and $t$ variable.
\item Set the ellipse centre point $(h, k)$ by fixing $h$ somewhere along the open line segment connecting the midpoints of the diagonals of the quadrilateral and hence determine $k$ from the equation of a line, for example:
\begin{eqnarray}
h =  \frac{1}{2}\left(\frac{s}{2}+\frac{A}{2}\right), \quad
k = \left(h-\frac{s}{2}\right)\left(\frac{t-B-C}{s-A}\right) + \frac{t}{2}
\end{eqnarray}
\item To solve for the ellipse tangent to the four sides of the quadrilateral, we can solve for the ellipse tangent to the three sides of a triangle whose vertices are the complex points
\begin{eqnarray}
z_{1} = 0, \quad
z_{2} = A+Bi, \quad
z_{3} = -\frac{At-Bs}{s-A}i
\end{eqnarray}
and the two ellipse foci  are then the zeroes of the equation
\begin{eqnarray}
p_{h}(z)&=&(s-A)z^{2}-2(s-A)(h-ik)z-(B-iA)(s-2h)C
\end{eqnarray}
whose discriminant can be denoted by $r(h)=r_{1}(h)+ir_{2}(h)$ where
\begin{align} \nonumber
r_1 \;=\; &4 \left(\left(s-A\right)^{2}-\left(t-B-C\right)^{2}\right)\left(\frac{h-A}{2}\right)^{2} \\ \nonumber
 &+4 \left(s-A\right)\left(A\left(s-A\right)+B\left(B-t\right)+C\left(C-t\right)\right)\left(\frac{h-A}{2}\right) \\
 &+ \left(s-A\right)^{2}\left(A^{2}-\left(C-B\right)^{2}\right) \\ \nonumber
r_2 \;=\; &8\left(t-B-C\right)\left(s-A\right)\left(\frac{h-A}{2}\right)^{2} \\ \nonumber
&+ 4\left(s-A\right)\left(At+Cs+Bs-2AB\right)\left(\frac{h-A}{2}\right) \\
&+ 2A\left(s-A\right)^{2}\left(B-C\right)
\end{align}Thus we need to determine the quartic polynomial $u(h)=|r(h)|^{2}={r_1(h)}^{2}+{r_2(h)}^{2}$ and we can then solve for the ellipse semimajor axis, $a$, and semiminor axis, $b$, from the equations
\begin{eqnarray}
a^{2}-b^{2} \;=\; \sqrt{ \frac{1}{\left(16\left(s-A\right)^{4}\right)}u(h)} 
\end{eqnarray}
\begin{eqnarray}
a^{2}b^{2} \;=\; \frac{1}{4}\left(\frac{C}{\left(s-A\right)^{2}}\right)\left(2\left(Bs-A\left(t-C\right)\right)h - ACs\right)\left(2h-A\right)\left(2h-s\right) 
\end{eqnarray}
by parameterising $R=a^{2}-b^{2}$ and $W=a^{2}b^{2}$ to obtain
\begin{eqnarray}
a \;=\; \sqrt{ \frac{1}{2}\left(\sqrt{R^{2}+4W}+R\right)}, \quad
b \;=\; \sqrt{ \frac{1}{2}\left(\sqrt{R^{2}+4W}-R\right)}
\end{eqnarray}
\item Knowing the axes we can generate the ellipse and float its tilt angle $\delta$ until it sits tangent to each side of the quadrilateral, using the inclined ellipse equation
\begin{eqnarray}
\rho^{2} \;=\; \frac{a^{2}b^{2}}{\left(\frac{a^{2}+b^{2}}{2}\right)-\left(\frac{a^{2}-b^{2}}{2}\right)\cos\left(2\omega'-2\delta\right)}
\end{eqnarray}
where $\omega'=\omega+\delta$ and $\omega$ is the angle from the semimajor axis to a radial line $\rho$ on the ellipse.
\end{enumerate}
\noindent
\newline
\textbf{Drag modelling.} The evolution of CMEs as they propagate from the Sun through the heliosphere is a complex process, simplified by using a parameterised drag model. Comparing equation~(\ref{drag}) and equation~(\ref{pdrag}): 
\begin{eqnarray}
	\label{drag3}
	\alpha r^{-\beta}&=&\frac{1}{2}\frac{A_{cme} C_{D} \rho_{sw}}{ M_{cme} }
\end{eqnarray}
where $C_{D}$ and $M_{cme}$ are approximately constant, and $A_{cme}$ and $\rho_{sw}$ are functions of distance expected to have a power-law form. We can therefore represent their combined behaviour as a single power law, as in equation~(\ref{drag3}). For example, if we assume a density profile of $\rho_{sw}(r)=\rho_{0}r^{-2}$, and a cylindrical CME of area $A_{cme}(r)=A_{0}r$, then from equation~(\ref{drag3}) we expect $\beta=1$. The $\alpha$ parameter, representative of the strength of the interaction, is then determined by the constants $A_{0}$, $M_{cme}$ and $C_{D}$, such that high mass, small volume, CMEs are less affected by drag than low mass, large volume, CMEs. This method of parameterisation has been shown to reproduce the kinematic profiles of a large number of events$^{44}$. We assume an additional parameter, $\gamma$, to indicate the type of drag, suggested to be either linear ($\gamma=1$) or quadratic ($\gamma=2$). While this parameterisation may obscure some of the complex interplay between the various quantities, it does not affect the most crucial part that we are trying to test: is aerodynamic drag an appropriate model and, if so, which regime (linear or quadratic) best characterises the kinematics.

A bootstrapping technique$^{60}$ was used to obtain statistically significant parameter ranges from the drag model of equation~(\ref{pdrag}). This technique involves the following steps:
\begin{enumerate}
\item An initial fit to the data $y$ is obtained, yielding the model fit $\hat{y}$ with parameters $\vec{p}$.
\item The residuals of the fit are calculated: $\epsilon = y - \hat{y}$.
\item The residuals are randomly resampled to give $\epsilon^{*}$.
\item The model is then fit to a new data vector $y^{*} = y + \epsilon^{*}$ and the parameters $\vec{p}$ stored.
\item Steps 3--4 are repeated many times (10,000).
\item Confidence intervals on the parameters are determined from the resulting distributions.
\end{enumerate}
In our case the model parameters were: the initial height $h_{cme}$ of the CME at the start of the modelling; the speed $v_{sw}$ of the solar wind at 1\,AU; the velocity $v_{cme}$ of the CME at the start of the modelling; and the drag parameters $\alpha$, $\beta$, and $\gamma$. In order to test for self-consistency we allowed the observationally known parameters of initial CME height and velocity to vary in the bootstrapping procedure, and recovered comparable values. The parameters $\alpha$ and $\beta$ were in reasonable agreement with values from previous studies$^{44}$.
\newline
\subsection*{References}

\begin{enumerate}

\item[1.] Schwenn, R., Dal Lago, A., Huttunen, E., Gonzalez, W. D. The association of coronal mass ejections with their effects near the Earth. {\it Ann. Geophys.} {\bf 23,} 1033--1059 (2005).
\item[2.] Prang{\'e}, R. {\it et al.} An interplanetary shock traced by planetary auroral storms from the Sun to Saturn. {\it Nature} {\bf 432,} 78--81 (2004). 
\item[3.] {\it Severe Space Weather Events -- Understanding Societal and Economic Impacts Workshop Report, National Research Council.} (National Academies Press, 2008).
\item[4.] Chen, J. Theory of prominence eruption and propagation: interplanetary consequences. {\it J. Geophys. Res.} {\bf 101,} 27499--27520 (1996).
\item[5.] Antiochos, S. K., DeVore, C. R., Klimchuk, J. A. A model for solar coronal mass ejections. {\it Astrophys. J.} {\bf 510,} 485--493 (1999).
\item[6.] Kliem, B., T{\"o}r{\"o}k, T. Torus instability. {\it Phys. Rev. Lett.} {\bf 96,} 255002 (2006).
\item[7.] Moore, R. L., Sterling, A. C. in {\it Solar Eruptions and Energetic Particles}, Gopalswamy, N., Mewaldt, R., Torsti, J. Eds. (American Geophysical Union, Washington, DC, 2006), {\bf165,} 43--57.
\item[8.] Xie, H. {\it et al.}, On the origin, 3D structure and dynamic evolution of CMEs near solar minimum. {\it Sol. Phys.} {\bf 259,} 143--161 (2009). 
\item[9.] Kilpua, E. K. J. {\it et al.} STEREO observations of interplanetary coronal mass ejections and prominence deflection during solar minimum period. {\it Ann. Geophys.} {\bf 27,} 4491--4503 (2009).
\item[10.] Chan{\'e}, E., Jacobs, C., van der Holst, B., Poedts, S., Kimpe, D. On the effect of the initial magnetic polarity and of the background wind on the evolution of CME shocks. {\it Astron. Astrophys.} {\bf 432,} 331--339 (2005).
\item[11.] Filippov, B. P., Gopalswamy, N., Lozhechkin, A. V. Non-radial motion of eruptive filaments. {\it Sol. Phys.} {\bf 203,} 119--130 (2001).
\item[12.] Gopalswamy, N., Dal Lago, A., Yashiro, S., Akiyama, S. The expansion and radial speeds of coronal mass ejections. {\it Cen. Eur. Astrophys. Bull.} {\bf 33,} 115--124 (2009).
\item[13.] Riley, P., Crooker, N. U. Kinematic treatment of coronal mass ejection evolution in the solar wind. {\it Astrophys. J.} {\bf 600,} 1035--1042 (2004).
\item[14.] Odstr{\v c}il, D., Pizzo, V. J. Three-dimensional propagation of coronal mass ejections in a structured solar wind flow 2. CME launched adjacent to the streamer belt. {\it J. Geophys. Res.} {\bf 104,} 493--504 (1999).
\item[15.] Maloney, S. A., Gallagher, P. T., McAteer, R. T. J. Reconstructing the 3-D trajectories of CMEs in the inner Heliosphere. {\it Sol. Phys.} {\bf 256,} 149--166 (2009).
\item[16.] Gonz{\'a}lez-Esparza, J. A., Lara, A., P{\'e}rez-Tijerina, E., Santill{\'a}n, A., Gopalswamy, N. A numerical study on the acceleration and transit time of coronal mass ejections in the interplanetary medium. {\it J. Geophys. Res. (Space Physics)} {\bf 108,} 1039 (2003).
\item[17.] Tappin, S. J. The deceleration of an interplanetary transient from the Sun to 5 AU. {\it Sol. Phys.} {\bf 233,} 233--248 (2006).
\item[18.] Cargill, P. J. On the aerodynamic drag force acting on interplanetary coronal mass ejections. {\it Sol. Phys.} {\bf 221,} 135--149 (2004).
\item[19.] Cremades, H., Bothmer, V. On the three-dimensional configurations of coronal mass ejections. {\it Astron. Astrophys.} {\bf 422,} 307--322 (2004).
\item[20.] Zhao, X. P., Plunkett, S. P., Liu, W. Determination of geometrical and kinematical properties of halo coronal mass ejections using the cone model. {\it J. Geophys. Res.} {\bf 107,} 1223 (2002).
\item[21.] Xie, H., Ofman, L., Lawrence, G. Cone model for halo CMEs: application to space weather forecasting. {\it J. Geophys. Res.} {\bf 109,} 3109 (2004).
\item[22.] D{\'e}moulin, P., Nakwacki, M. S., Dasso, S., Mandrini, C. H. Expected in situ velocities from a hierarchical model for expanding interplanetary coronal mass ejections. {\it Sol. Phys.} {\bf 250,} 347--374 (2008).
\item[23.] Howard, T. A., Nandy, D., Koepke, A. C. Kinematics properties of solar coronal mass ejections: correction for projection effects in spacecraft coronagraph measurements. {\it J. Geophys. Res. (Space Physics)} {\bf 113,} 1104 (2008).
\item[24.] Moran, T. G., Davila, J. M. Three-dimensional polarimetric imaging of coronal mass ejections. {\it Science} {\bf 305,} 66--71 (2004).
\item[25.] Kaiser, M. L. {\it et al.} The STEREO mission: an introduction. {\it Space Sci. Rev.} {\bf 136,} 5--16 (2008).
\item[26.] Mierla, M. {\it et al.} On the 3-D reconstruction of coronal mass ejections using coronagraph data. {\it Ann. Geophys.} {\bf 28,} 203--215 (2010).
\item[27.] Inhester, B. Stereoscopy basics for the STEREO mission. (arXiv: astro-ph/0612649, 2006).
\item[28.] Aschwanden, M. J., W{\"u}lser, J. P., Nitta, N. V., Lemen, J. R. First three-dimensional reconstruction of coronal loops with the STEREO A and B spacecraft. I. Geometry. {\it Astrophys. J.} {\bf 679,} 827--842 (2008).
\item[29.] Howard, R. A. {\it et al.} Sun earth connection coronal and heliospheric investigation (SECCHI). {\it Space Sci. Rev.} {\bf 136,} 67--115 (2008).
\item[30.] Liewer, P. C. {\it et al.} Stereoscopic analysis of the 19 May 2007 erupting filament. {\it Sol. Phys.} {\bf 256,} 57--72 (2009).
\item[31.] Srivastava, N., Inhester, B., Mierla, M., Podlipnik, B. 3D reconstruction of the leading edge of the 20 May 2007 partial halo CME. {\it Sol. Phys.} {\bf 259,} 213--225 (2009).
\item[32.] Wood, B. E., Howard, R. A., Thernisien, A., Plunkett, S. P., Socker, D. G. Reconstructing the 3D morphology of the 17 May 2008 CME. {\it Sol. Phys.} {\bf 259,} 163--178 (2009).
\item[33.] Howard, T. A., Tappin, S. J. Three-dimensional reconstruction of two solar coronal mass ejections using the STEREO spacecraft. {\it Sol. Phys.} {\bf 252,} 373--383 (2008).
\item[34.] Davis, C. J. {\it et al.}, Stereoscopic imaging of an Earth-impacting solar coronal mass ejection: a major milestone for the STEREO mission. {\it Geophys. Res. Lett.} {\bf 36,} 8102 (2009).
\item[35.] Davis, C. J., Kennedy, J., Davies, J. A. Assessing the accuracy of CME speed and trajectory estimates from STEREO observations through a comparison of independent methods. {\it Sol. Phys.} {\bf 263,} 209--222 (2010).
\item[36.] Liu, Y. {\it et al.} Geometric triangulations of imaging observations to track coronal mass ejections continuously out to 1 AU. {\it Astrophys. J. Lett.} {\bf 710,} L82--L87 (2010).
\item[37.] Thernisien, A., Vourlidas, A., Howard, R. A. Forward modeling of coronal mass ejections using STEREO/SECCHI data. {\it Sol. Phys.} {\bf 256,} 111--130 (2009).
\item[38.] Byrne, J. P., Gallagher, P. T., McAteer, R. T. J., Young, C. A. The kinematics of coronal mass ejections using multiscale methods. {\it Astron. Astrophys.} {\bf 495,} 325--334 (2009).
\item[39.] Gopalswamy, N. {\it et al.} Prominence eruptions and coronal mass ejection: a statistical study using microwave observations. {\it Astrophys. J.} {\bf 586,} 562--578 (2003).
\item[40.] Pneuman, G. W., Kopp, R. A. Gas-magnetic field interactions in the solar corona. {\it Sol. Phys.} {\bf 18,} 258--270 (1971).
\item[41.] Banaszkiewicz, M., Axford, W. T., McKenzie, J. F. An analytic solar magnetic field model. {\it Astron. Astrophys.} {\bf 337,} 940--944 (1998).
\item[42.] Arge, C. N., Pizzo, V. J. Improvement in the prediction of solar wind conditions using near-real time solar magnetic field updates. {\it J. Geophys. Res.} {\bf 105,} 10465--10480 (2000).
 \item[43.] Vr{\v s}nak, B., Gopalswamy, N. Influence of the aerodynamic drag on the motion of interplanetary ejecta. {\it J. Geophys. Res.} {\bf 107,} 1019 (2002).
\item[44.] Vr{\v s}nak, B. Deceleration of coronal mass ejections. {\it Sol. Phys.} {\bf 202,} 173--189 (2001).
\item[45.] Schrijver, C. J., Elmore, C., Kliem, B., T{\" o}r{\" o}k, T., Title, A. M. Observations and modelling of the early acceleration phase of erupting filaments involved in coronal mass ejections. {\it Astrophys. J.} {\bf 674,} 586--595 (2008).
\item[46.] Lin, C.-H., Gallagher, P. T., Raftery, C. L. Investigating the driving mechanisms of coronal mass ejections. {\it Astron. Astrophys.} in press (2010).
\item[47.] Bothmer, V., Schwenn, R. Eruptive prominences as sources of magnetic clouds in the solar wind. {\it Space Sci. Rev.} {\bf 70,} 215--220 (1994).
\item[48.] Berdichevsky, D. B., Lepping, R. P., Farrugia, C. J. Geometric considerations of the evolution of magnetic flux ropes. {\it Phys. Rev. E} {\bf 67,} 036405 (2003).
\item[49.] Cargill, P. J., Schmidt, J., Spicer, D. S., Zalesak, S. T. Magnetic structure of overexpanding coronal mass ejections: numerical models. {\it J. Geophys. Res.} {\bf 105,} 7509--7520 (2000).
\item[50.] Yashiro, S. {\it et al.} A catalog of white light coronal mass ejections observed by the SOHO spacecraft. {\it J. Geophys. Res. (Space Physics)} {\bf 109,} 7105 (2004).
\item[51.] MacQueen, R. M., Hundhausen, A. J., Conover, C. W. The propagation of coronal mass ejection transients. {\it J. Geophys. Res.} {\bf 91,} 31--38 (1986).
\item[52.] Vr{\v s}nak, B., Vrbanec, D., {\v C}alogovi{\'c}, J., {\v Z}ic, T. The role of aerodynamic drag in dynamics of coronal mass ejections. {\it IAU Symposium}, {\bf 257,} 271--277 (2009).
\item[53.] McComas, D. J. {\it et al.} Solar Probe Plus: Report of the Science and Technology Definition Team (STDT). NASA/TM--2008--214161, NASA, 2008.
\item[54.] Hassler, D. {\it et al.} Solar Orbiter: Exploring the Sun-heliosphere Connection. ESA/SRE(2009)5, 2009.
\item[55.] Young, C. A., Gallagher, P. T. Multiscale edge detection in the corona. {\it Sol. Phys.} {\bf 248,} 457--469 (2008).
\item[56.] Pizzo, V. J., Biesecker, D. A. Geometric localization of STEREO CMEs. {\it Geophys. Res. Lett.} {\bf 31,} 21802 (2004).
\item[57.] deKoning, C. A., Pizzo, V. J., Biesecker, D. A. Geometric localization of CMEs in 3D space using STEREO beacon data: first results. {\it Sol. Phys.} {\bf 256,} 167--181 (2009).
\item[58.] Horwitz, A. Finding ellipses and hyperbolas tangent to two, three, or four given lines. {\it Southwest J. Pure Appl. Math.} {\bf 1,} 6--32 (2002).
\item[59.] Horwitz, A. Ellipses of maximal area and of minimal eccentricity inscribed in a convex quadrilateral. {\it Austral. J. Math. Anal. Appl.} {\bf 2,} 1--12 (2005).
\item[60.] Efron, B., Tibshirani, R. J. {\it An Introduction To The Bootstrap.} (Chapman \& Hall/CRC, 1993).

\end{enumerate}

\subsection*{Acknowledgements}

This work is supported by the Science Foundation Ireland under Grants No. 07-RFP-PHYF399 and 729S0DAZ. R.T.J.M.A. was a Marie Curie Fellow at TCD. The STEREO/SECCHI project is an international consortium of the Naval Research Laboratory (USA), Lockheed Martin Solar and Astrophysics Lab (USA), NASA Goddard Space Flight Center (USA), Rutherford Appleton Laboratory (UK), University of Birmingham (UK), Max-Planck-Institut f\"{u}r Sonnen-systemforschung (Germany), Centre Spatial de Liege (Belgium), Institut d'Optique Th\'{e}orique et Appliqu\'{e}e (France), and Institut d'Astrophysique Spatiale (France). Simulation results have been provided by the Community Coordinated Modeling Center at Goddard Space Flight Center through their public Runs on Request system (http://ccmc.gsfc.nasa.gov). The CCMC is a multi-agency partnership between NASA, AFMC, AFOSR, AFRL, AFWA, NOAA, NSF and ONR. The ENLIL with Cone Model was developed by D. Odstrcil at the University of Colorado at Boulder. We acknowledge the use of WIND data.

\subsection*{Author Contributions}

J.P.B. developed the method and performed the analysis. S.A.M carried out the drag modelling and bootstrapping procedure and contributed to the analysis. J.M.R. developed the visualisation suite. R.T.J.M.A. guided the data prepping. P.T.G. supervised the research. J.P.B., S.A.M., R.T.J.M.A., and P.T.G. discussed the results and implications and contributed to the manuscript at all stages.

\subsection*{Additional information}

{\bf Supplementary Information} accompanies this paper on \\ http://www.nature.com/ncomms/journal/v1/n6/full/ncomms1077.html
\newline
{\bf Competing financial interests:} The authors declare no competing financial interests.
\newline
{\bf Reprints and permission} information is available online at \\
http://www.nature.com/naturecommunications
\newline
{\bf How to cite this article:} Byrne, J.P. {\it et al.} Propagation of an Earth-directed coronal mass ejection in three dimensions. {\it Nat. Commun.} 1:74 doi: 10.1038/ncomms1077 (2010).
\newline
\newline
{\bf Figure 1: Composite of STEREO-A and B images taken by the SECCHI instruments of the 12 Dec. 2008 CME.} Panel {\bf a} indicates the STEREO spacecraft locations, separated by an angle of 86.7$^{\circ}$ at the time of the event. Panel {\bf b} shows the prominence eruption observed in EUVI-B off the north-west limb from approximately 03:00~UT which is considered to be the inner material of the CME. The multiscale edge detection and corresponding ellipse characterisation are overplotted in COR1. Panel {\bf c} shows that the CME is Earth-directed, being observed off the east limb in STEREO-A and the west limb in STEREO-B.
\newline
\newline
{\bf Figure 2: The epipolar geometry used to constrain the reconstruction of the 12 Dec. 2008 CME front.} The reconstruction is performed using an elliptical tie-pointing technique within epipolar planes containing the two STEREO spacecraft$^{27}$. For example, one of any number of planes will intersect the ellipse characterisation of the CME at two points in each image from STEREO-A and B. Panel {\bf a} illustrates how the resulting four sight-lines intersect in 3D space to define a quadrilateral that constrains the CME front in that plane$^{56,57}$. Inscribing an ellipse within the quadrilateral such that it is tangent to each sight-line$^{58,59}$ provides a slice through the CME that matches the observations from each spacecraft. Panel {\bf b} illustrates how a full reconstruction is achieved by stacking multiple ellipses from the epipolar slices. Since the positions and curvatures of these inscribed ellipses are constrained by the characterised curvature of the CME fronts in the stereoscopic image pair, the modelled CME front is considered an optimum reconstruction of the true CME front. Panel {\bf c} illustrates how this is repeated for every frame of the eruption to build the reconstruction as a function of time and view the changes to the CME front as it propagates in 3D. While the ellipse characterisation applies to both the leading edges and, when observable, the flanks of the CME, only the outermost part of the reconstructed front is shown here for clarity, and illustrated in Supplementary Movie 2.
\newline
\newline
{\bf Figure 3: Kinematic and morphological properties of the 3D reconstruction of the 12 Dec. 2008 CME front.} Panel {\bf a} shows the velocity of the middle of the CME front with corresponding drag model and, inset, the early acceleration peak. Measurement uncertainties are indicated by one standard deviation error-bars. Panel {\bf b} shows the declinations from the ecliptic (0$^{\circ}$) of an angular spread across the front between the CME flanks with a power-law fit indicative of non-radial propagation. It should be noted that the positions of the flanks are subject to large scatter: as the CME enters each field-of-view the location of a tangent to its flanks is prone to moving back along the reconstruction in cases where the epipolar slices completely constrain the flanks. Hence the `Midtop/Midbottom of Front' measurements better convey the southward dominated expansion. Panel {\bf c} shows the angular width of the CME with a power-law expansion. For each instrument the first three points of angular width measurement were removed since the CME was still predominantly obscured by each instrument's occulter.
\newline
\newline
{\bf Figure 4: The in-situ solar wind plasma and magnetic field measurements obtained using instruments on the WIND spacecraft.} From top to bottom the panels show proton density, bulk flow speed, proton temperature, and magnetic field strength and components. The red dashed lines indicate the predicted window of CME arrival time from our ENLIL with Cone Model run (08:09\,--\,13:20~UT on 16 Dec. 2008). We observe a magnetic cloud flux rope signature behind the front, highlighted by the blue dash-dotted lines.
\newline
\newline
{\bf Figure 5: An ellipse inscribed within a convex quadrilateral.} An isometry of the plane is applied such that the quadrilateral has vertices $(0,0)$, $(A,B)$, $(0,C)$, $(s,t)$. The ellipse has center $(h,k)$, semimajor axis $a$, semiminor axis $b$, tilt angle $\delta$, and is tangent to each side of the quadrilateral.

\begin{figure}[!p]
\centerline{\includegraphics[width=\linewidth]{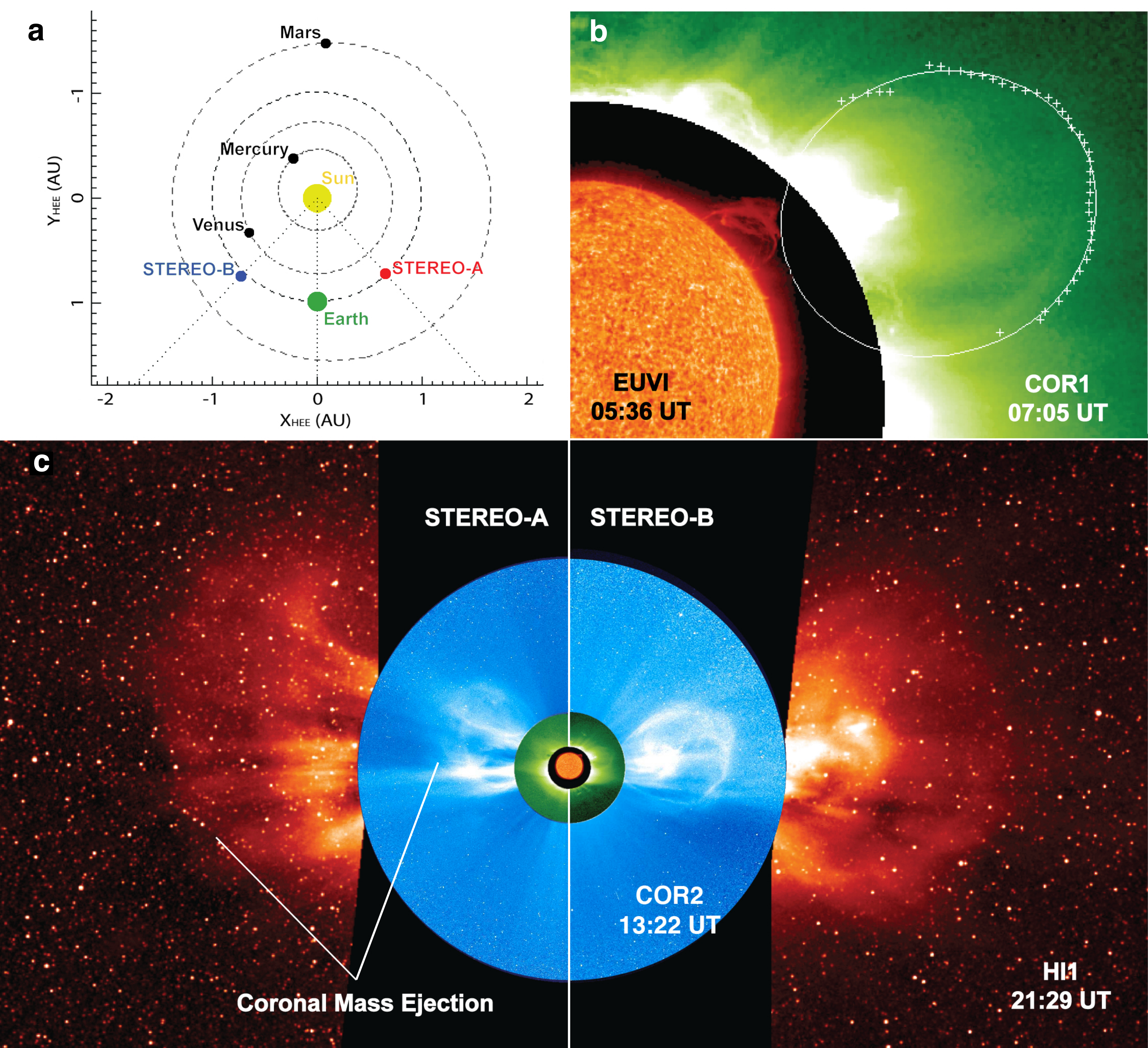}}
\end{figure}

\begin{figure}[!p]
\centerline{\includegraphics[angle=90, width=\linewidth]{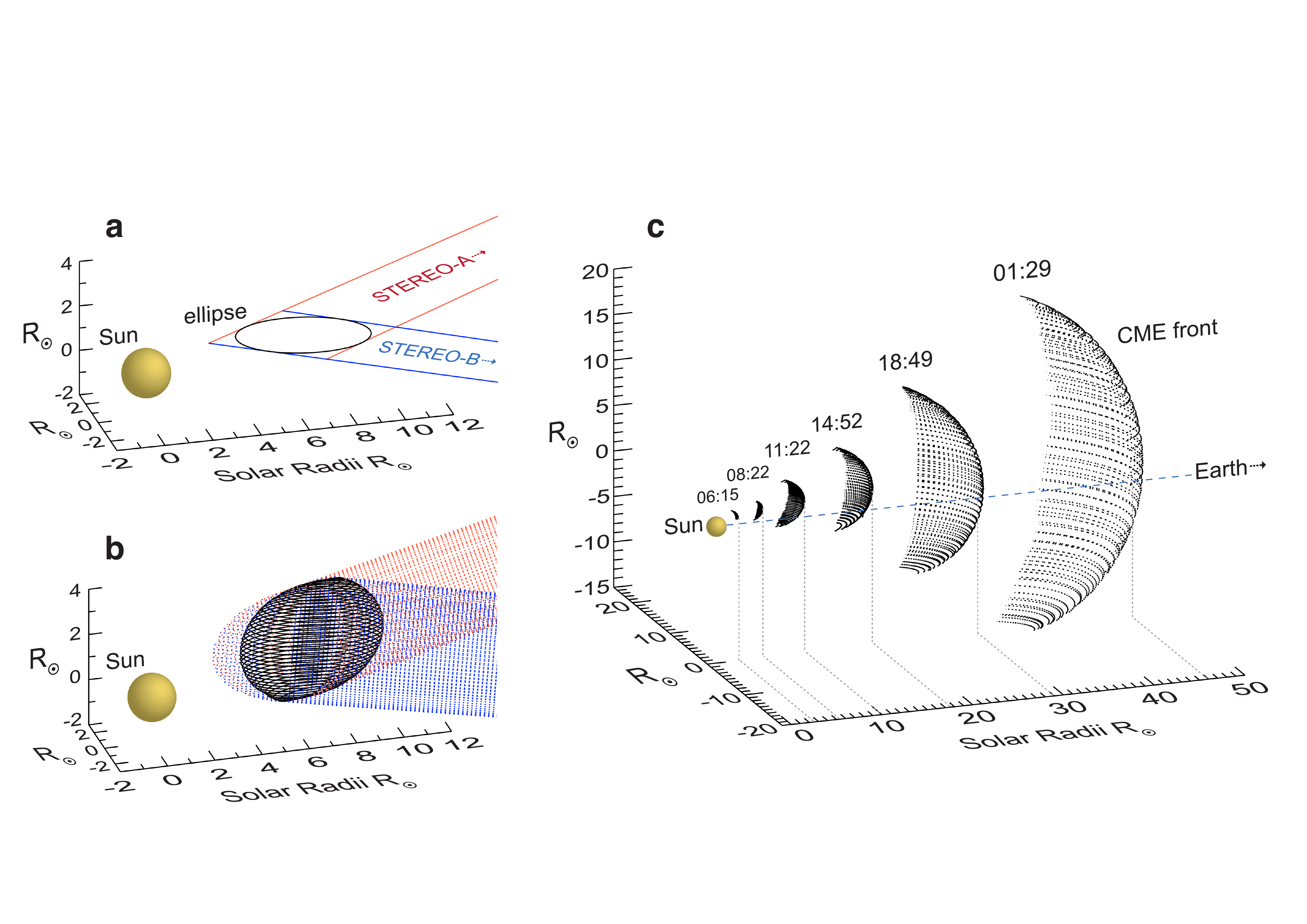}}
\end{figure}

\begin{figure}[!p]
\centerline{\includegraphics[width=\linewidth]{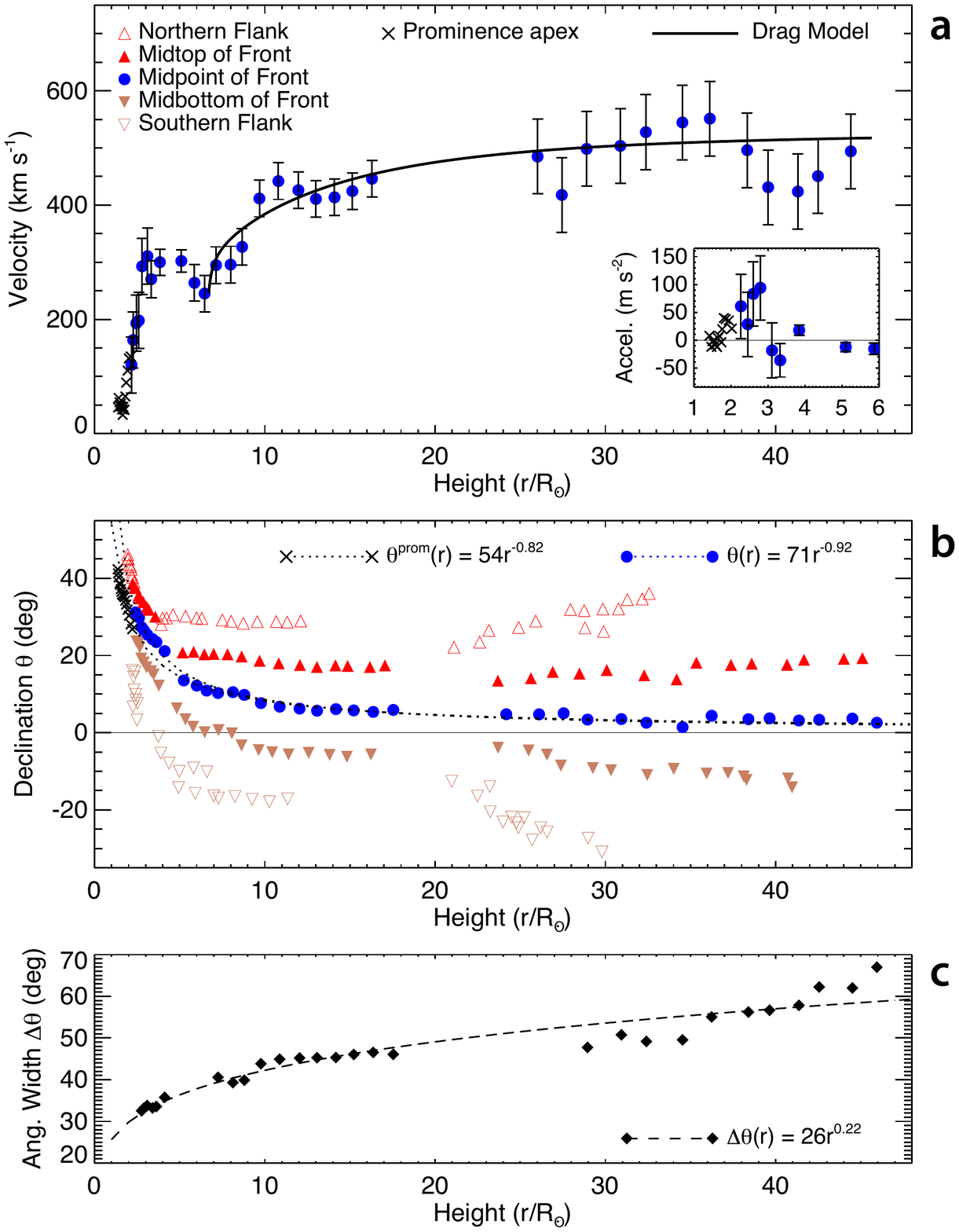}}
\end{figure}

\begin{figure}[!p]
\centerline{\includegraphics[width=\linewidth]{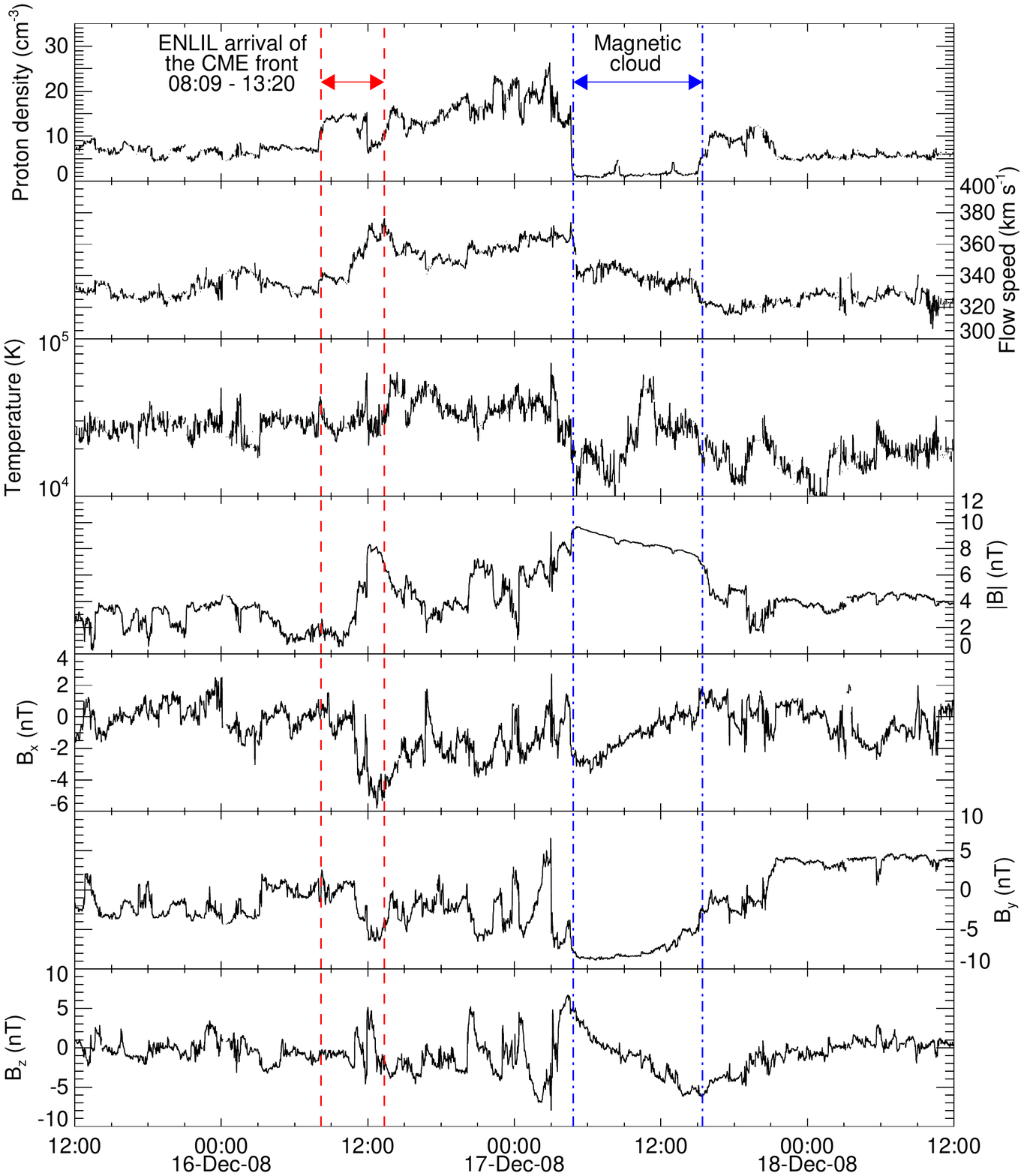}}
\end{figure}

\begin{figure}[!p]
\centerline{\includegraphics[width=\linewidth]{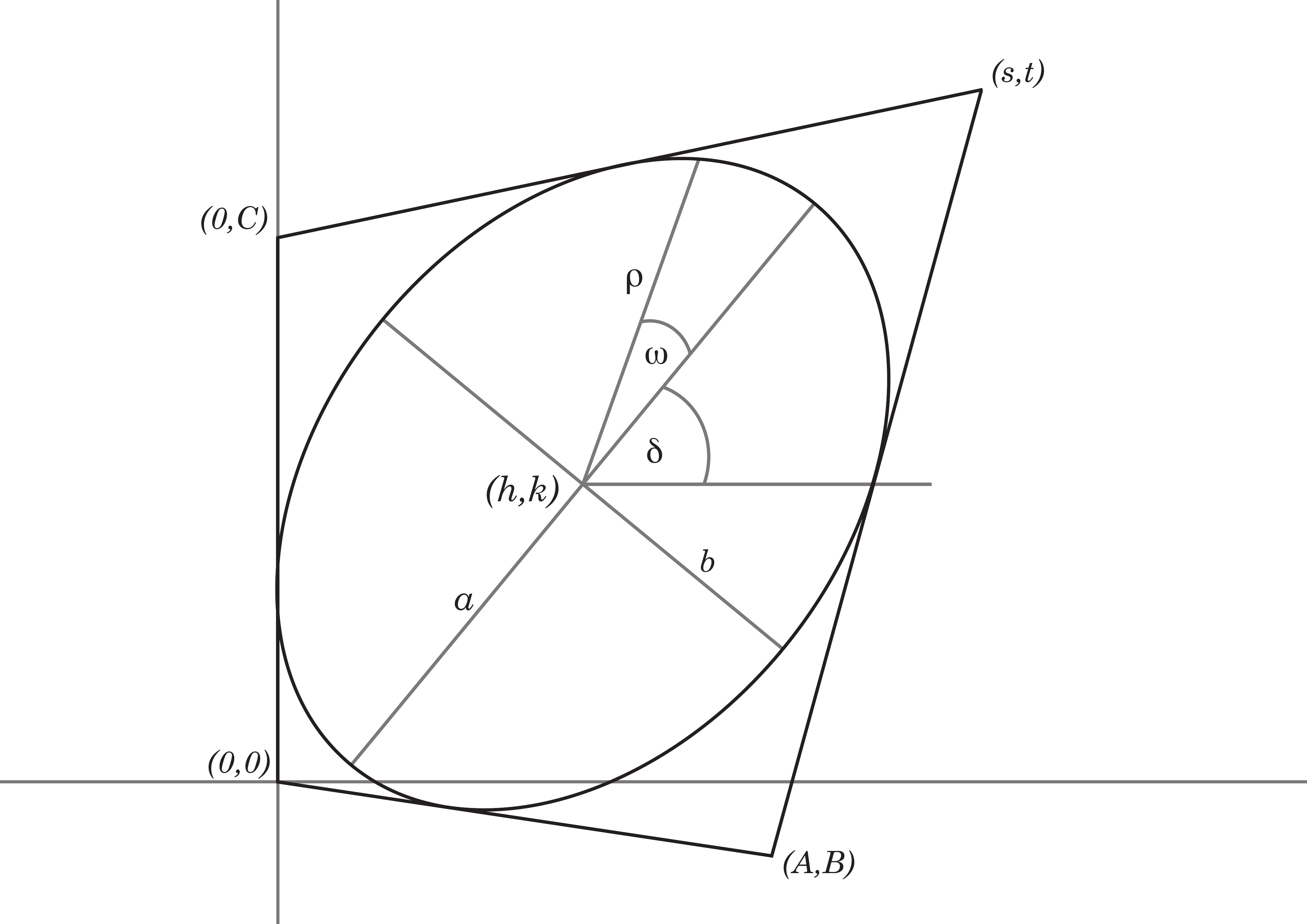}}
\end{figure}

\end{document}